\documentclass[twocolumn]{aastex61}
\pdfoutput=1 %for arXiv submission
\usepackage{amsmath,amstext}
\usepackage[T1]{fontenc}
\usepackage{apjfonts} 
\usepackage{multirow}
\usepackage[figure,figure*]{hypcap}

\newcommand{\photu}{photon units}
\newcommand{\galex}{{\it GALEX}}
\newcommand{\iras}{{\it IRAS}}
\newcommand{\planck}{{\it Planck}}

\shorttitle{Diffuse Background at High Galactic Latitudes}
\shortauthors{Akshaya et al.}

\begin{document}

\title{The Diffuse Radiation Field at High Galactic Latitudes}

\author{M. S. Akshaya}
\affiliation{Department of Physics, Christ, Bengaluru 560 029, India}
\email{akshaya.subbanna@gmail.com}
\author{Jayant Murthy}
\affiliation{Indian Institute of Astrophysics, Bengaluru 560 034, India}
\email{jmurthy@yahoo.com}
\author{S. Ravichandran}
\affiliation{Department of Physics, Christ, Bengaluru 560 029, India}
\email{ravichandran.s@christuniversity.in}
\author{R. C. Henry}
\affiliation{Henry A. Rowland Department of Physics and Astronomy, The Johns Hopkins University, Baltimore, MD 21218, USA}
\email{henry@jhu.edu}
\author{James Overduin}
\affiliation{Department of Physics, Astronomy and Geosciences, Towson University, Towson, MD 21252, USA}
\email{joverduin@towson.edu}

\begin{abstract}
We have used \galex\ observations of the North and South Galactic poles to study the diffuse ultraviolet background at locations where the Galactic light is expected to be at a minimum. We find offsets of 230 -- 290 \photu\ in the FUV (1531 \AA) and 480 -- 580 \photu\ in the NUV (2361 \AA). Of this, approximately 120 \photu\ can be ascribed to dust scattered light and another 110 (190 in the NUV) \photu\ to extragalactic radiation. The remaining radiation is, as yet, unidentified and amounts to 120 -- 180 \photu\ in the FUV and 300 -- 400 \photu\ in the NUV. We find that molecular hydrogen fluorescence contributes to the FUV when the 100 \micron\ surface brightness is greater than 1.08 MJy sr$^{-1}$.

\end{abstract}

\keywords{dust --- local interstellar matter --- surveys --- ultraviolet: general --- ultraviolet: ISM}

\section{Introduction}
The diffuse radiation at high latitudes is, by definition, a combination of the diffuse Galactic light (DGL) and the extragalactic background light (EBL). The largest component of the DGL at low latitudes is the light from Galactic plane stars scattered by interstellar dust \citep{Jura1979} but this will be at a minimum at the poles where there is little dust. Thus much of the diffuse light at the poles might be expected to be from the EBL \citep{Bowyer1991, Henry1991}. As a result, there were many observations of the cosmic ultraviolet background at the pole and we have listed them in Table \ref{tab:pole_obs}. The typical surface brightness was 200 -- 300 ph\ cm$^{-2}$ s$^{-1}$ sr$^{-1}$ \AA$^{-1}$ (hereafter photon units) in the far ultraviolet (FUV: 1300 -- 1800 \AA) and 300 -- 600 \photu\ in the near ultraviolet (NUV: 1800 -- 3200 \AA). 

The EBL is comprised of several parts with the most significant being the integrated light of galaxies which \citet{Driver2016} found to be 60 -- 81 \photu\ (FUV) and 121 -- 181 \photu\ (NUV). These values are model-dependent but differ by no more than about 20 \photu\ \citep{Xu2005,Voyer2011,Gardner2000}. There may be smaller contributions from the integrated light of QSOs (16 -- 30 \photu: \citet{Madau1992}) and the IGM (\textless\ 20 \photu: \citet{Martin1991}) for a total EBL of 96 -- 131 \photu\ in the FUV and 157 -- 231 \photu\ in the NUV. Phenomenological models of the cosmic spectral energy distribution are increasingly consistent with observational data and semi-analytic models, except in the ultraviolet, where they differ by as much as 100 photon units \citep[Fig. 9]{Andrews2018}. A good review of the current state of uncertainty in ultraviolet EBL intensity may be found in \citet[Fig. 5]{Hill2018}.

\citet{Henry2015} has argued strongly that there is an additional component to the DGL, unrelated to dust-scattered starlight. Much of the evidence for this component comes from \galex\ observations of the Galactic poles in the FUV from \citet{Murthy2010}. We have used an improved reduction of the diffuse background \citep{Murthy2014apj} with a Monte Carlo model for the dust scattered light \citep{Murthy2016} to further explore the background in the vicinity of both Galactic poles in the far-ultraviolet (FUV: 1531 \AA) and the near-ultraviolet (NUV: 2361 \AA). 

\begin{table}[!h]
\caption{Polar Observations}
\begin{tabular}{lcc}
\hline
References & Wavelength  & Offset \\
           & (\AA)            & (photon units) \\
\hline
\hline
\citet{Anderson1979} & 1230 -- 1680 & 285 $\pm$ 32 \\
\citet{Paresce1979} & 1350 -- 1550 & 300 $\pm$ 60 \\
\citet{Paresce1980} & 1350 -- 1550 & \textless300 \\
\citet{Joubert1983} & 1690 & 300 -- 690 \\
 & 2200 & 160 -- 360 \\
\citet{Jakobsen1984} & 1590 & \textless550 \\
 & 1710 & \textless900 \\
 & 2135 & \textless1300 \\
\citet{Tennyson1988} & 1800 -- 1900 & 300 $\pm$ 100 \\
& 1900 -- 2800 & 400 $\pm$ 200 \\
\citet{Onaka1991} & 1500 & 200 -- 300 \\
\citet{Feldman1981} & 1200 -- 1670 & 150 $\pm$ 50 \\
\citet{Henry1993} & 1500  & 300 $\pm$ 100 \\
\citet{Murthy1995} & 1250 -- 2000 & 100 -- 400 \\
\citet{Hamden2013} & 1344 -- 1786 & 300 \\
\citet{Boissier2015} & 1528 & 315 \\
\citet{Murthy2016} & 1531 & 300 \\
& 2361 & 600 \\

\hline
\end{tabular}
\label{tab:pole_obs}
\end{table}

\section{Data}
The \galex\ mission \citep{Martin2005, Morrissey2007} took observations covering most of the sky in the FUV and NUV bands. An observation consisted of one or more visits with exposure times of 100 -- 1000 seconds each which could be added together to reach total integration times of as high as 100,000 seconds. The original data from the mission were distributed as FITS (Flexible Image Transport System) files with a pixel size of $1.5''$. \citet{Murthy2014apj} masked out the stars, rebinned to $2'$ pixels and subtracted the foreground emission \citep{Murthy2014apss} to produce a map of the diffuse background over the sky. We have used the visit-level data from \citet{Murthy2014apj}, available from the High Level Science Products (HLSP) data repository\footnote{https://archive.stsci.edu/prepds/uv-bkgd/} at the Space Telescope Science Institute, to study the diffuse emission at the Galactic poles.

\begin{figure}
\includegraphics[scale=0.35]{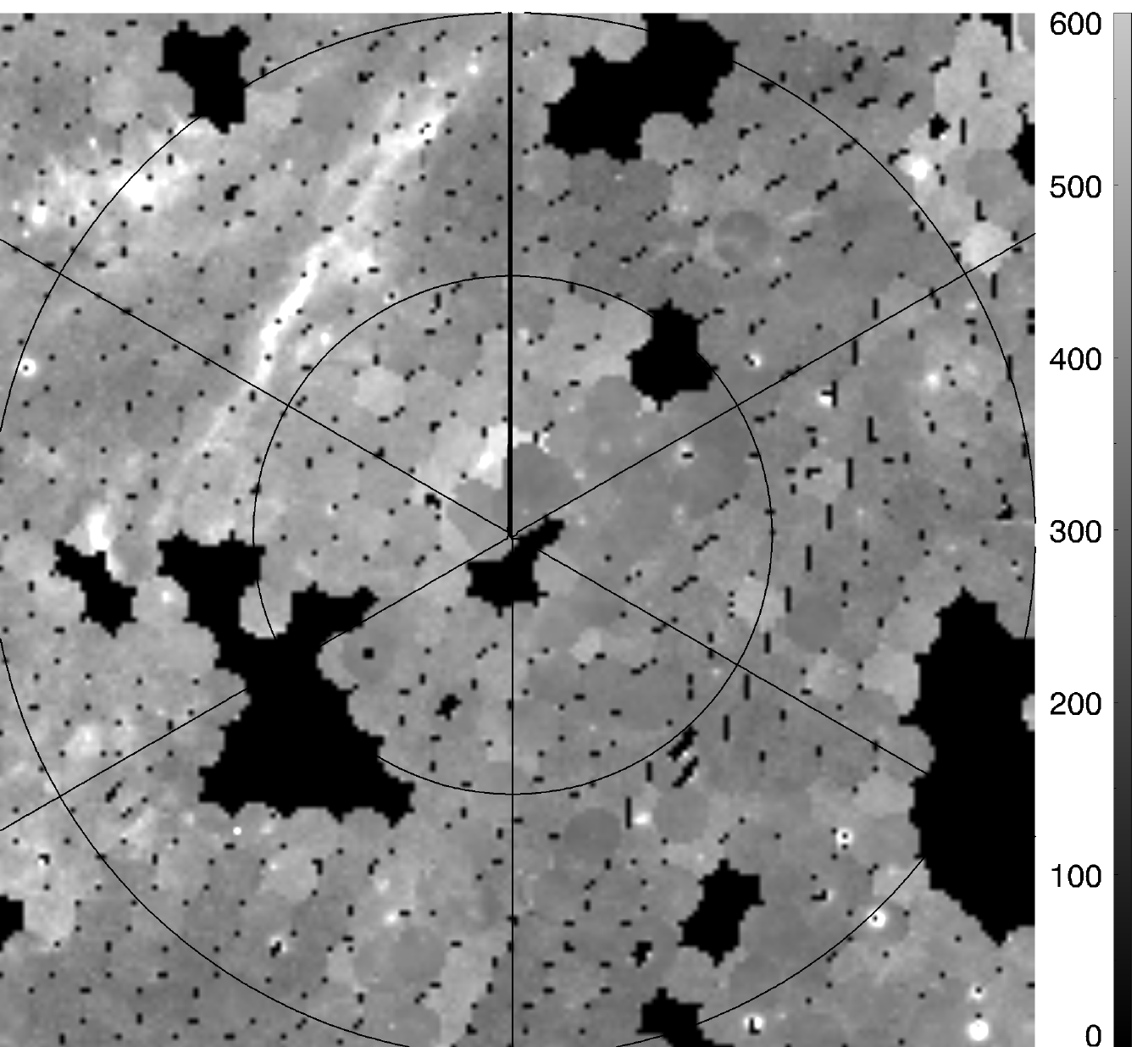}
\includegraphics[scale=0.35]{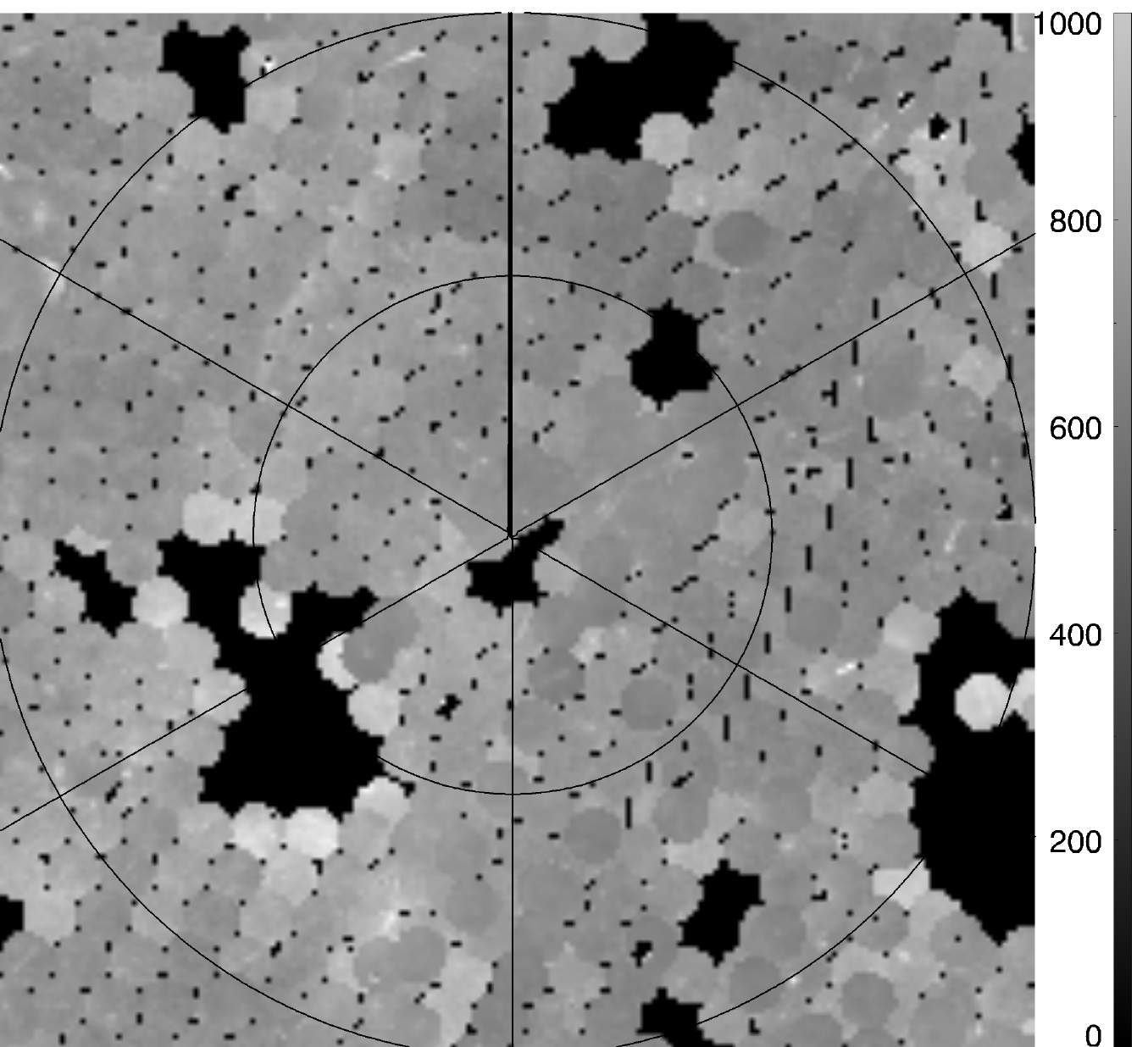}
\includegraphics[scale=0.35]{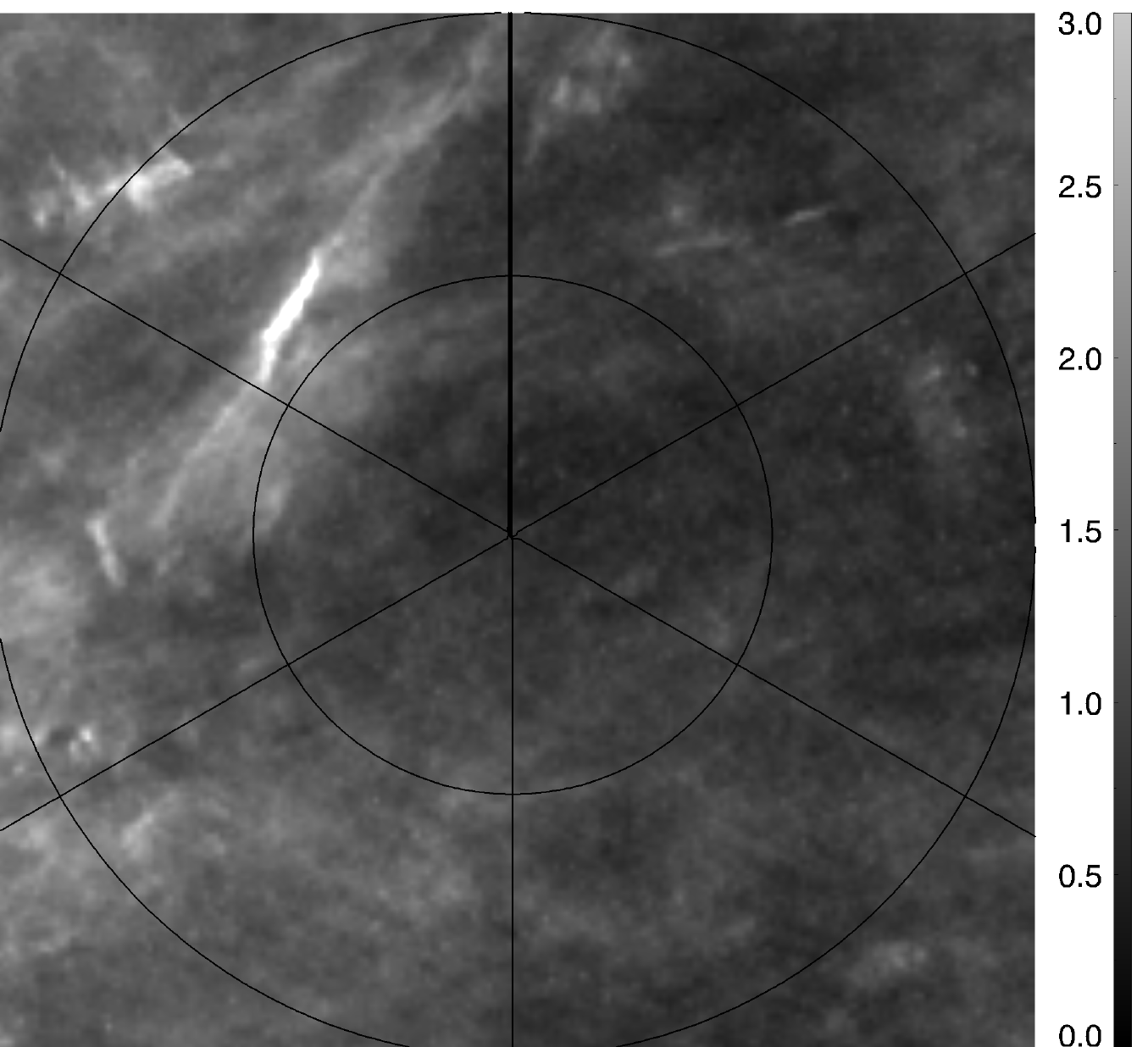}
\caption{Observed surface brightness in FUV (top) and NUV (middle) from \galex\ and 100 \micron\ from \cite{Schlegel1998} (bottom). The FUV and NUV maps are in \photu\ and the 100 \micron\ map is in MJy sr$^{-1}$. Black areas were not observed by \galex. The NGP is at the center with lines of latitude at $80^{\circ}$ and $85^{\circ}$ and lines of longitude every $60^{\circ}$ starting from $0^{\circ}$ at the top increasing clockwise. Bright spots in the FUV image are due to artifacts around bright stars and were not included in the analysis.}
\label{fig:ngp_images}
\end{figure}

\begin{figure}
\includegraphics[scale=0.35]{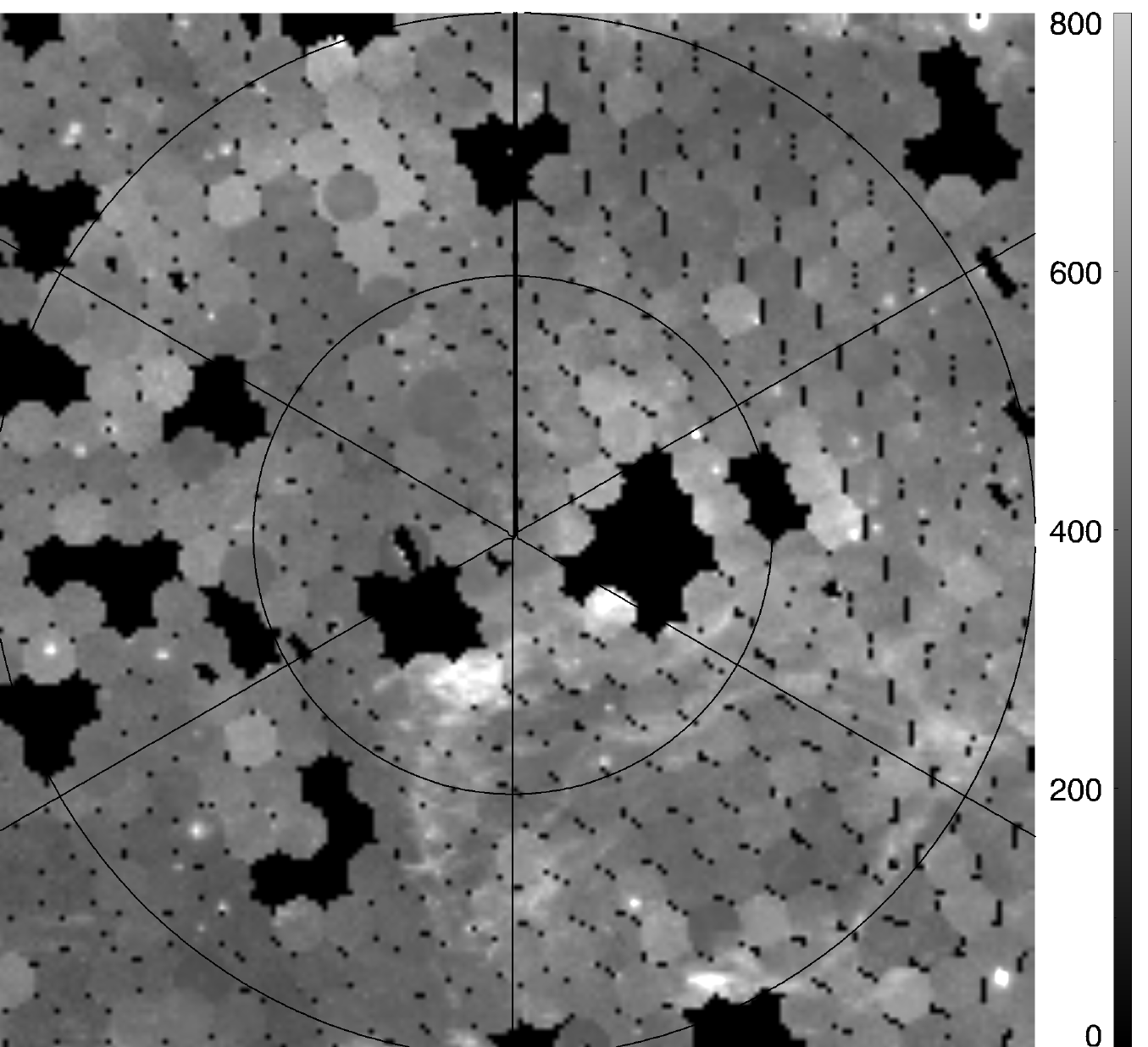}
\includegraphics[scale=0.35]{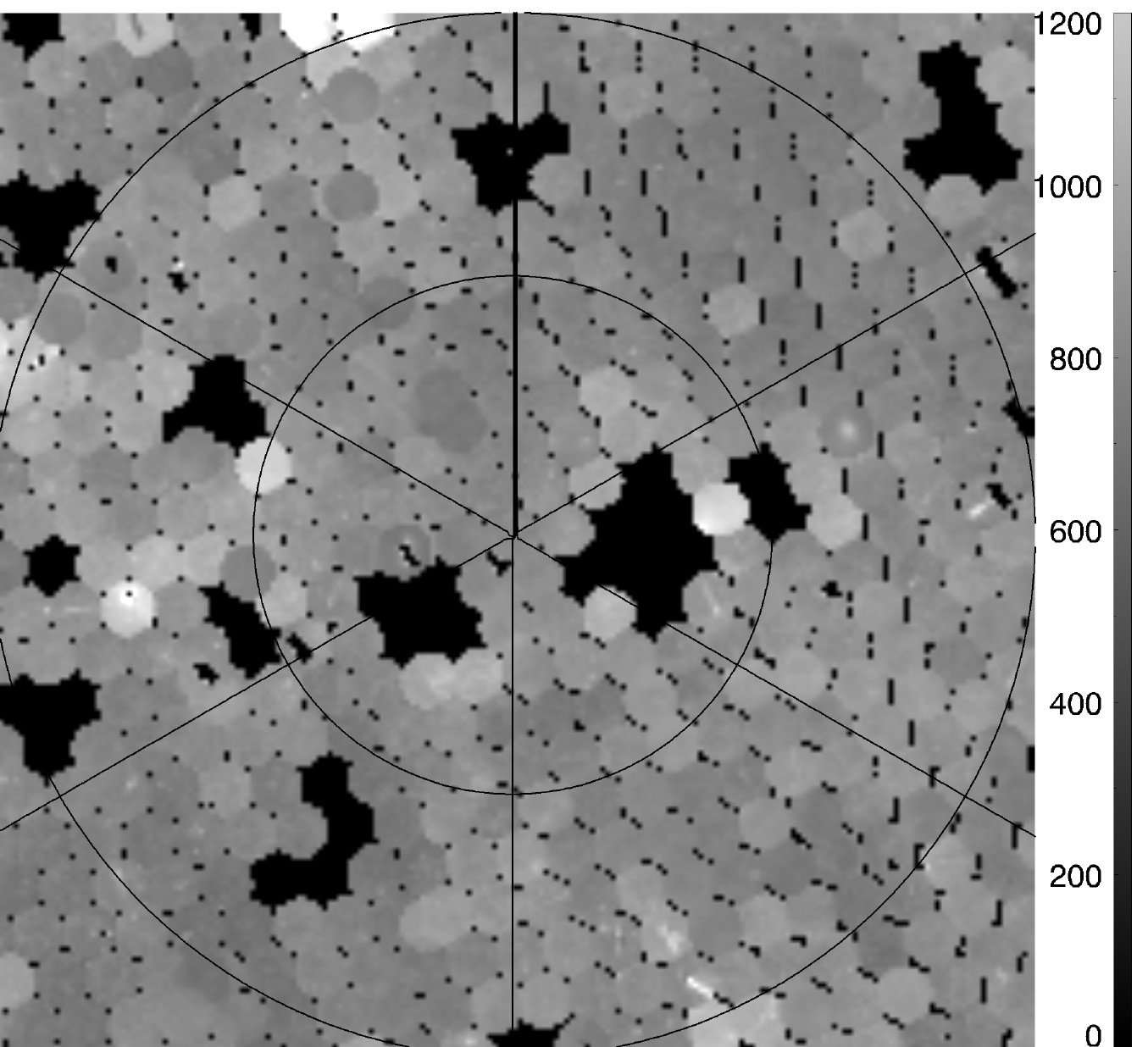}
\includegraphics[scale=0.35]{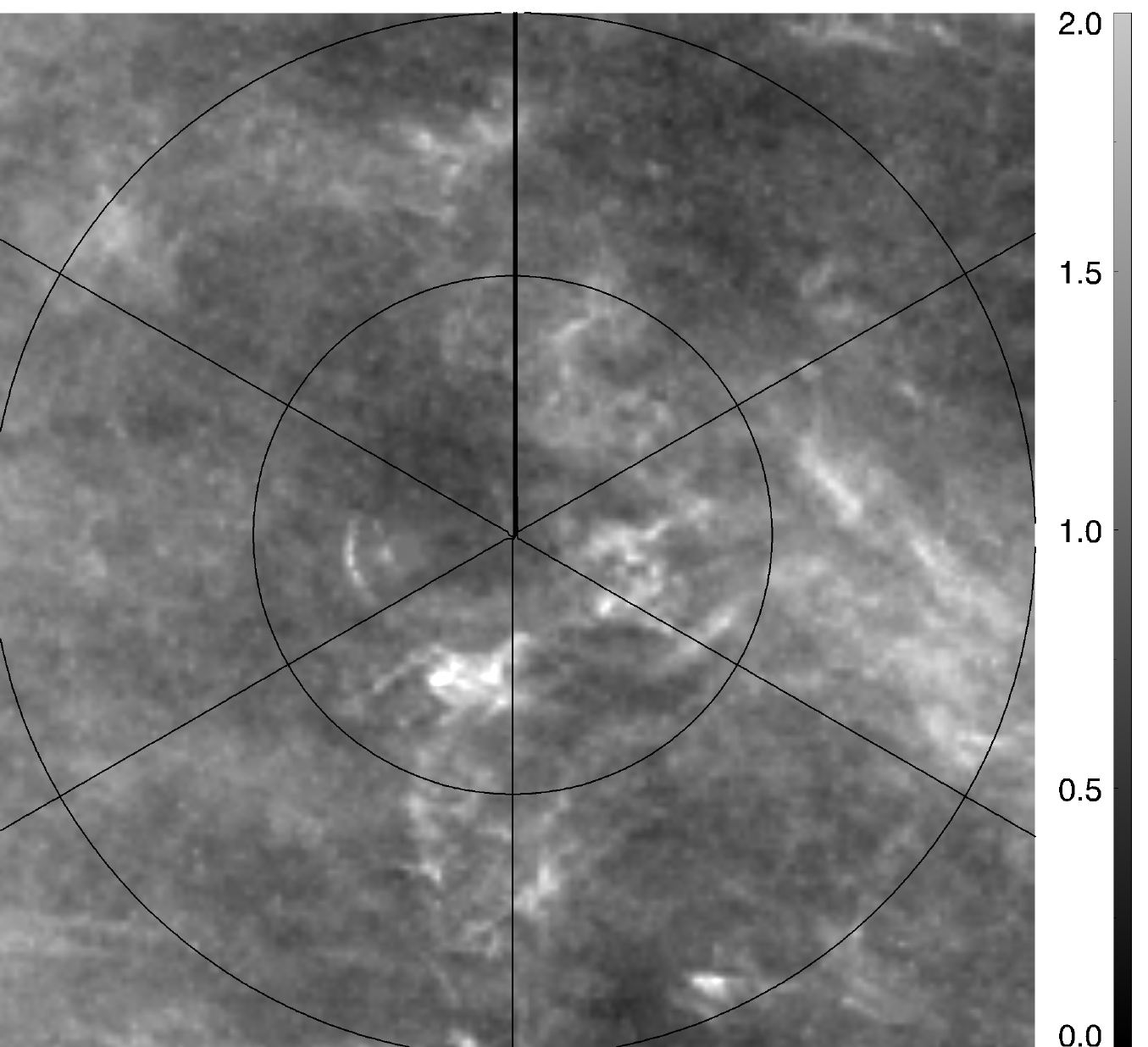}
\caption{Same as in Fig. \ref{fig:ngp_images} but for the SGP. The SGP is at the center with lines of latitude at $-80^{\circ}$ and $-85^{\circ}$ and lines of longitude every $60^{\circ}$ starting from $0^{\circ}$ at the top increasing anti-clockwise.}
\label{fig:sgp_images}
\end{figure}

We further rebinned the original $2'$ bins of \citet{Murthy2014apj} by a factor of 3 (into $6'$ bins) to improve the signal-to-noise and the resultant maps are shown for the  North Galactic pole (NGP) in Fig. \ref{fig:ngp_images} and the South Galactic pole (SGP) in  Fig. \ref{fig:sgp_images} along with the 100 \micron\ maps from \cite{Schlegel1998}, also rebinned to $6'$ pixels. Although one might expect a good correlation between the FUV and the NUV and between both UV bands and the IR \citep{Hamden2013,Murthy2014apj}, there is much less structure in the NUV image than in the 100 \micron\ images or, indeed, in the FUV. 

Given that these are archival data, the number of visits and the exposure times per field fluctuate wildly but with most of the field observed in multiple visits. The deepest observation was the Subaru Deep Field \citep{Subaru2004}, which was targeted by \galex\ \citep{Ly2009} as part of the overall saturation coverage of that region by a number of different observatories. There were a total of 99 different visits in the FUV and 169 in the NUV with exposure times from 80 -- 1700 seconds for each visit. The cumulative exposure times over the three years from Apr. 2004 to May 2007 is 83,031 seconds in the FUV and 164,369 seconds in the NUV. 

\begin{figure}
\includegraphics[trim = 1.5cm 12cm 2cm 5cm, clip, scale=0.5]{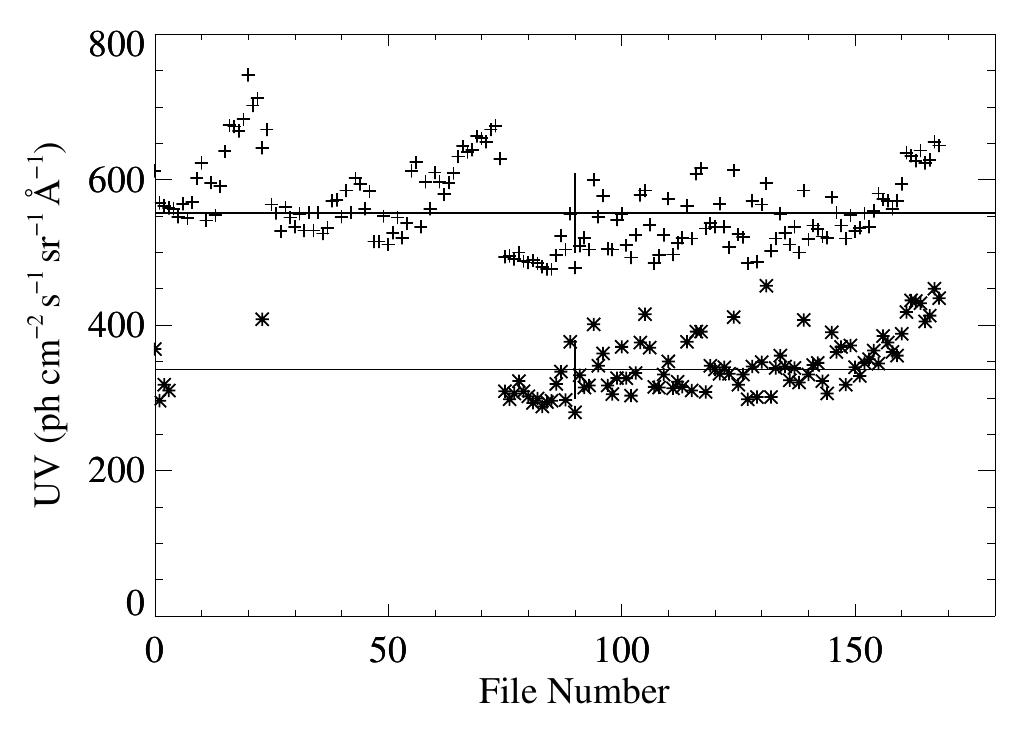}
\caption{Median values for FUV (*) and NUV (+) in each visit. The horizontal lines show the medians in each band over all visits with the standard deviation plotted as vertical lines in the center of the plot. Many FUV values are missing because there were no observations on those dates.}
\label{fig:visit_var}
\end{figure}

The primary source of uncertainty in the derived astrophysical background is the foreground emission (airglow in both bands and zodiacal light in the NUV), which is comparable to the astrophysical emission at high Galactic latitudes. We have tested the foreground subtraction by tracking the background surface brightness of a single $6'$ bin over all the visits in the Subaru field (Fig. \ref{fig:visit_var}). There are variations in both bands which, despite the missing FUV observations, are obviously correlated (r = 0.9). These are manifested as an increase in the overall background level of the image which we believe are due to changes in the radiation environment around the spacecraft but could not find any obvious trigger, either terrestrial or solar. The mean value of the background over all the visits in a $6'$ pixel is $346 \pm 41$ \photu\ in the FUV and $563 \pm 55$ \photu\ in the NUV and we have adopted these uncertainties in our analysis.

We took the individual visits and added them into polar grids (Fig. \ref{fig:ngp_images} and \ref{fig:sgp_images}), weighting each visit by its exposure time. Most of the field was covered by multiple visits and we assumed that the diffuse surface brightness in a given field was comprised of a constant DGL + EBL with any difference between visits being due to the uncharacterized foreground discussed above. We subtracted this difference from each visit, effectively setting the median level of the diffuse surface brightness to the minimum over all visits.

There is a bright point in the top of the NUV image of the SGP (Fig. \ref{fig:sgp_images}) due to nebulosity around the fifth magnitude star HD 224990 (B3V). We have not included those points in our analysis. Bright points in the FUV images are due to artifacts around hot stars and are not used in the analysis.

\section{Results}
\subsection{UV-IR Correlations}

\begin{figure}
\includegraphics[trim = 1.5cm 12cm 2cm 3.5cm, clip, scale=0.5]{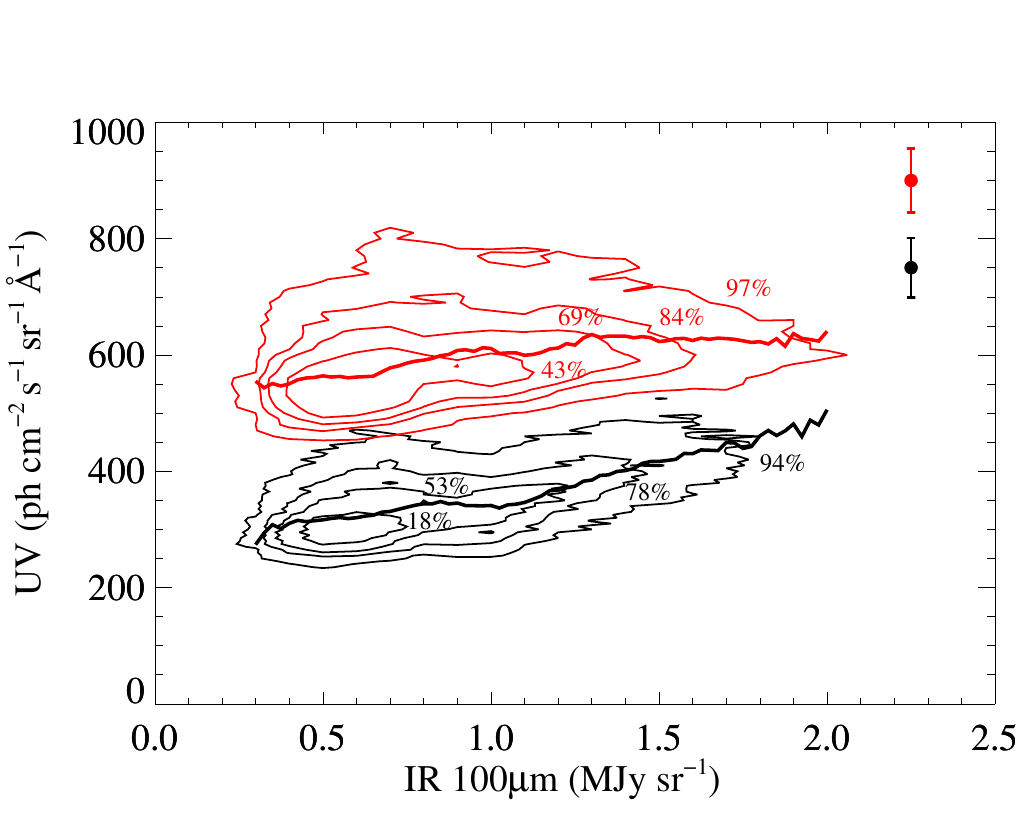}
\includegraphics[trim = 1.5cm 12cm 2cm 3.5cm, clip, scale=0.5]{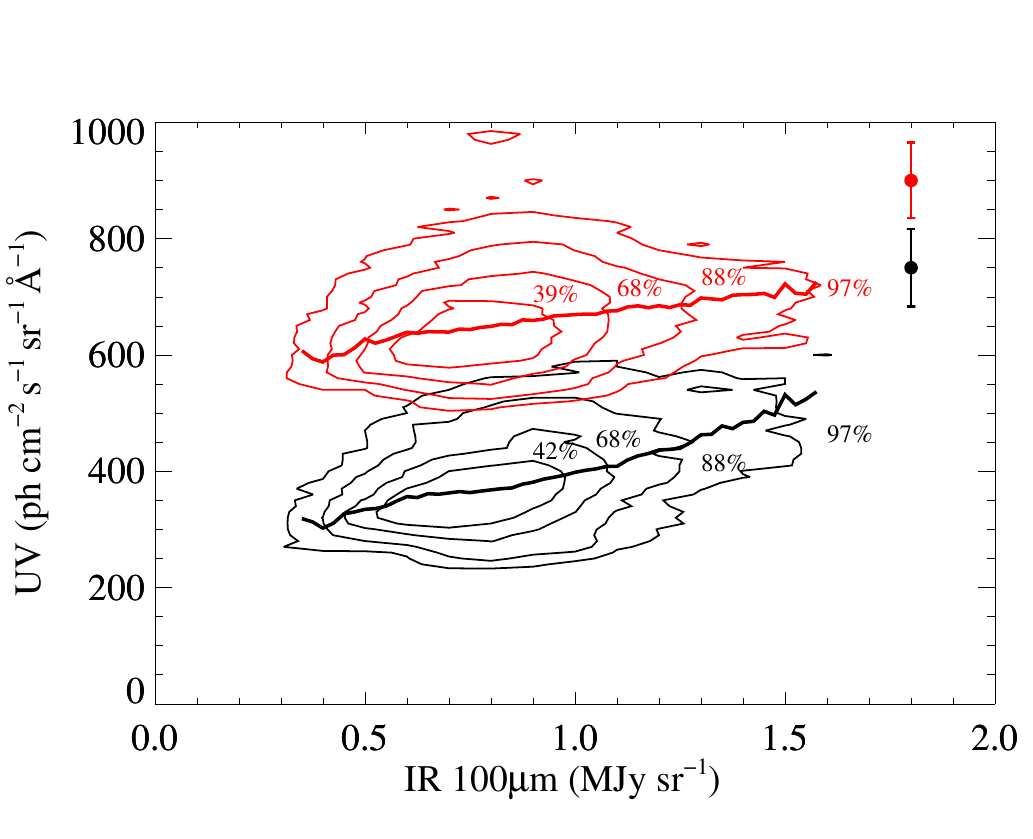}
\caption{Contour plots of the FUV (black contours) and NUV (red contours) at NGP (top) and SGP (bottom) where the IR bin size is 0.1 MJy sr$^{-1}$ and the UV bin size is 5 photon units. We have shown the mean surface brightness in the UV averaged over bins of 0.025 MJy sr$^{-1}$ in the IR. The error bars shown are representative of the standard deviation in the mean and are on the order of about 50 \photu\ in the NGP and 70 \photu\ in the SGP.}
\label{fig:uv_ir}
\end{figure}

Both the UV and the 100 \micron\ surface brightness track the presence of dust and should be linearly correlated at high Galactic latitudes where the optical depth is low. We have plotted the observed correlations in Fig. \ref{fig:uv_ir} and tabulated them in Table \ref{tab:ngp_corr}. The UV does indeed correlate with the IR but not as well as one might expect, as is apparent from a visual comparison of the images in Fig. \ref{fig:ngp_images} and \ref{fig:sgp_images}. The bright IR features such as Markkanen's Cloud \citep{Markkanen1979} are readily seen in the FUV at both poles but are not prominent in the NUV.

\begin{figure}
\includegraphics[trim = 1.5cm 12cm 2cm 3.5cm, clip, scale=0.5]{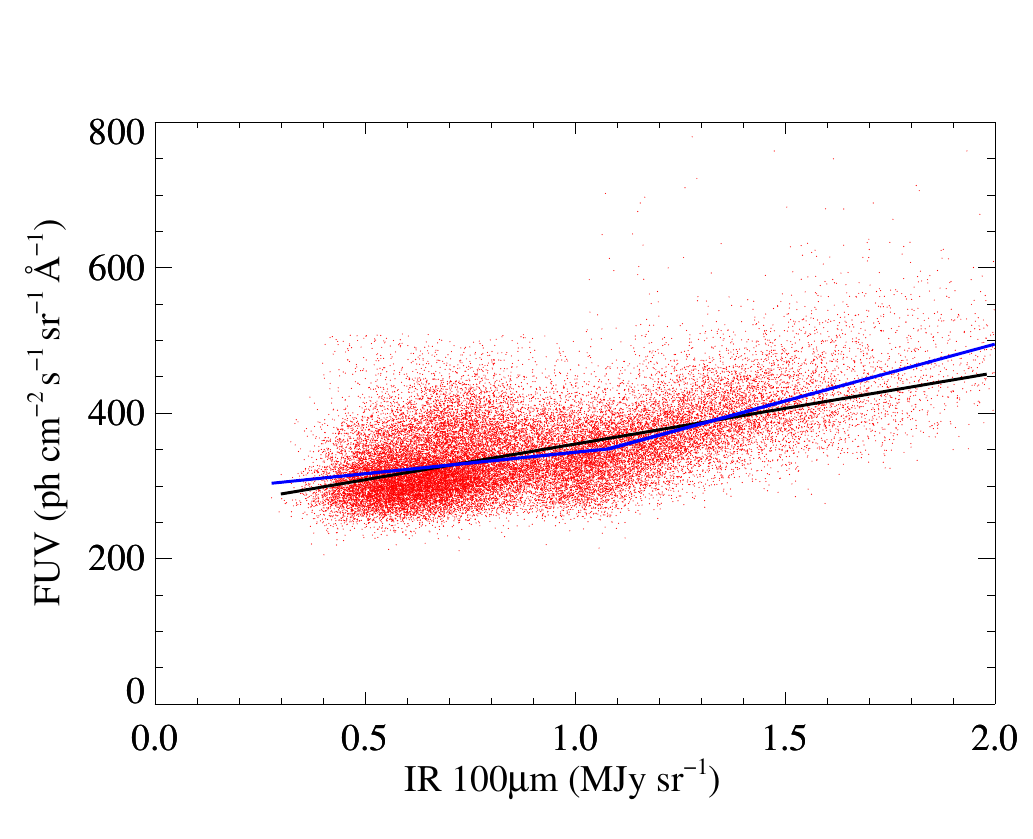}
\caption{FUV surface brightness plotted against IR surface brightness for the NGP. The black line shows a linear fit over all the data points and the blue line shows the best fit with the inflection point at 1.08 MJy sr$^{-1}$.}
\label{fig:fuv_inflection}
\end{figure}

\begin{table}
\centering 
\caption{Correlation coefficients}
\label{tab:ngp_corr}
\begin{tabular}{lllll}
\hline 
Bands         & p$^a$   & a$^b$  & b$^c$  & $\chi^2_\nu$ \\ \hline
\hline
\multicolumn{5}{c}{NGP}                              \\ \hline
FUV -- IRAS &      0.54 &       97.58 &       259.80 &       1.26 \\ 
NUV -- IRAS &      0.42 &       68 &       530.89 &       1.18 \\
FUV -- E(B-V) &      0.52 &       4245.40 &       250.11 &       1.24 \\
NUV -- E(B-V) &      0.40 &       2655.67 &       531.06 &       1.21 \\
\hline
\multicolumn{5}{c}{SGP}                               \\ \hline
FUV -- IRAS &      0.42 &       164.17 &       240.99 &       2.29 \\
NUV -- IRAS &      0.28 &       90.30 &       579.07 &       1.48 \\
FUV -- E(B-V) &      0.45 &       7967.55 &       211.85 &       2.18 \\
NUV -- E(B-V) &      0.29 &       4260.72 &       565.69 &       1.47 \\
\hline
\multicolumn{5}{c}{NGP (With inflection point)}\\
\hline
 FUV -- IRAS (< 1.08 MJy sr$^{-1}$)  &	0.27 &	57.43 &		288.27 & 	1.15 \\ 
 FUV -- IRAS (> 1.08 MJy sr$^{-1}$)  &  0.57 & 156.33 &     182.10 &	1.37 \\ 
\hline
\multicolumn{5}{l}{$^{a}$Spearman's correlation coefficient (P $\lll$ 0.05 for all the cases).}\\
\multicolumn{5}{l}{$^{b}$Scale factor }\\
\multicolumn{5}{l}{$^{c}$Offset (\photu)}\\ 
\end{tabular}
\end{table}

We noted an inflection point in the FUV/IR ratio in the NGP at an IR surface brightness of 1.08 MJy sr$^{-1}$ (Fig. \ref{fig:fuv_inflection}). We performed an F-Test \citep{Bevington2003} to investigate whether the additional term was justified and found an F-value of 1325 which is significant at greater than a 99.9\% level. \citet{Matsuoka2011}, perhaps coincidentally, found a similar inflection point at a 100 \micron\ surface brightness of 0.8 MJy sr$^{-1}$ in {\it Pioneer} optical data, which they identified with the cosmic infrared background (CIB:  \citet{Lagache2000}). In this scenario, both the CIB and the UV offset would represent that part of the background which is not correlated with interstellar dust. However, we would then expect a similar inflection point in the FUV data at the SGP or in the NUV at either pole which is not seen.

Another possibility is that the change in slope is due to molecular hydrogen (H$_{2}$) fluorescence \citep{Hurwitz1998} in the Werner bands kicking in at a 100 \micron\ surface brightness of 1.08 MJy sr$^{-1}$ ({logN${_H}$ = 20.2}). Canonically, H$_{2}$ is only formed at column densities greater than logN${_H}$ = 20.5 -- 20.7 \citep{Knapp1974, Savage1977, Franco1986, Reach1994}, when self-absorption protects the molecules from dissociation by ultraviolet photons. \citet{Jo2017} have found that the fraction of the total diffuse radiation in the form of fluorescent Werner band emission from molecular hydrogen is 5-10\% of the total observed surface brightness at the poles, or about 30 \photu. These observations were averaged over 10 -- 15 degrees at the poles and we find that the putative Werner band emission in the \galex\ data is about 60 \photu, not too far off from their observations. \cite{Gillmon2006} and \cite{Wakker2006} have found significant molecular gas at high latitudes at column densities of 20.2 < logN${_H}$ < 20.5, which \cite{Gillmon2006} attributed to a clumpy medium with the molecular gas concentrated in high density cirrus clouds. Unfortunately, we do not have the spectroscopic information needed to further investigate the emission and cannot further constrain the source of the rise in the FUV.

\subsection{Zero-Points}

\begin{table}[!h]
\centering
\caption{Components of the Background}
\begin{tabular}{lcc}
\hline
Component & FUV$^{a}$ & NUV$^{a}$\\
\hline
\hline
\multicolumn{3}{c}{NGP}\\
\hline
Observed & 288 $\pm$ 2 & 531 $\pm$ 2\\
EGL  & 114 $\pm$ 18 & 194 $\pm$ 37\\
Remainder & 174 $\pm$ 18 & 337 $\pm$ 37\\
\hline
\multicolumn{3}{c}{SGP}\\
\hline
Observed & 241 $\pm$ 2 & 579 $\pm$ 3\\
EGL  & 114 $\pm$ 18 & 194 $\pm$ 37 \\
Remainder & 127 $\pm$ 18 & 385$\pm$ 37\\
\hline
$^{a}$ \photu & & \\
\end{tabular}
\label{tab:back_limits}
\end{table}

The diffuse radiation at the poles is likely to be dominated by the EBL and the observed baseline will therefore place an upper limit on the EBL. The y intercepts for the FUV are 288 \photu\ in the NGP and 241 \photu\ in the SGP with the corresponding values for the NUV being 531 and 579 \photu\ in the NGP and SGP, respectively. Taken at face value, these are upper limits for the EBL and match well with earlier determinations of the background at the poles (Table \ref{tab:pole_obs}), including with \galex\ results from \citet{Hamden2013}, \citet{Boissier2015}, and \citet{Murthy2016}. As an independent check, we have calculated the slopes and offsets using the E(B - V) from \citet{Planck_thermal_dust2014} finding very similar offsets (Table \ref{tab:ngp_corr}). However, as discussed in the Introduction, the expected limits on the EBL are 96 -- 131 \photu\ in the FUV and 157 -- 231 \photu\ the NUV, or about half the observed value in the FUV and about one third in the NUV. This offset has been noted before (Table \ref{tab:pole_obs}) but with no definite identification \citep{Henry2015}.

\subsection{Correlation with E(B - V)}

\begin{figure}
\includegraphics[trim = 1.5cm 12cm 2cm 3.5cm, clip, scale=0.5]{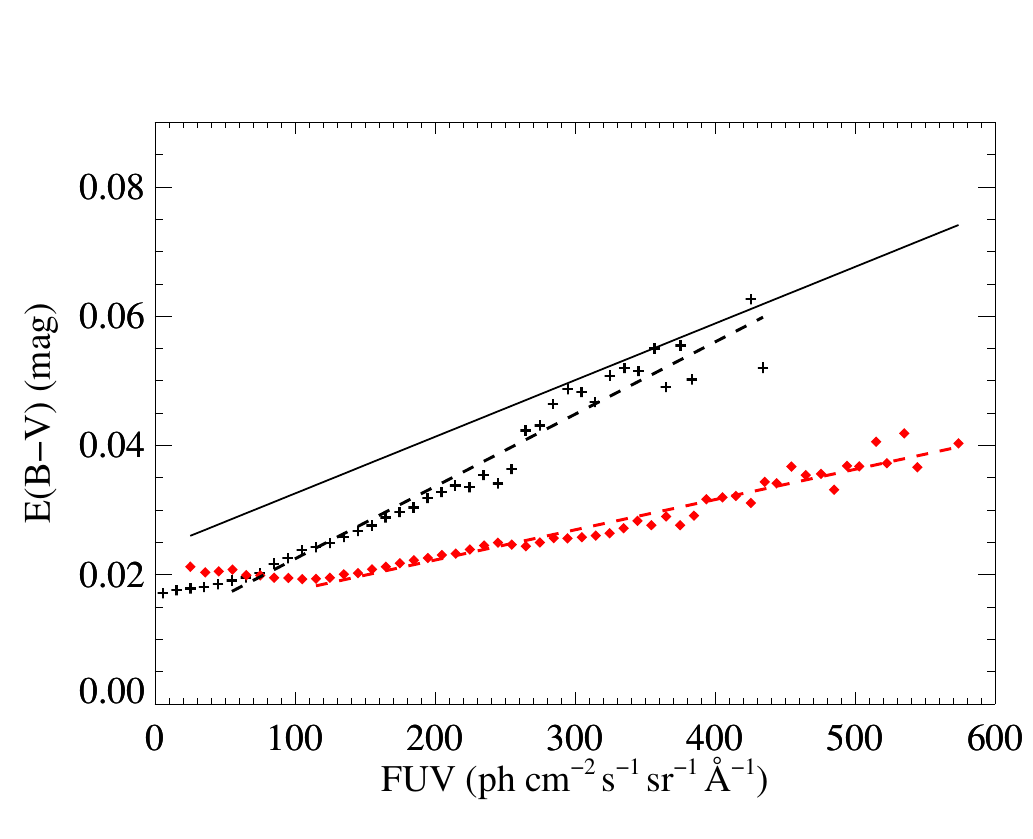}
\caption{E(B - V) from \citet{Planck_thermal_dust2014} plotted as a function of FUV for the NGP (plus signs) and the SGP (red diamonds), where the reddening has been averaged over the FUV bins. The straight line shows the relation derived by \citet{Boissier2015}. The dashed lines show our best fit to the reddening for the NGP (black line) and the SGP (red line). Note that, in each case, an offset has been subtracted from the FUV to account for the non-cirrus emission.}
\label{fig:fuvebv}
\end{figure}

Much of the \galex\ Ultraviolet Virgo Cluster Survey (GUVICS: \citet{Boissier2015}) falls within our area and our extracted diffuse values are in excellent agreement in the areas of overlap, despite independent approaches to the extraction of the diffuse radiation from the \galex\ observations. \citet{Boissier2015} subtracted what they termed as ``any emission not related to the cirrus'' from the EBL and from unknown Galactic sources, possibly including ``a very diffuse cirrus contribution'' and then derived a linear relationship between the FUV (in \photu ) and the reddening of E(B - V) = $0.02378 + 8.77\times 10^{-5} \times (FUV-315)$, where 315 \photu\ was their offset. They suggested that the diffuse UV background could be used to calculate the E(B - V) at a higher spatial resolution and precision than either the \iras\ data \citep{Schlegel1998} or the \planck\ data \citep{Planck_thermal_dust2014}. This method does indeed show promise and we have attempted the same with our data over both poles (Fig. \ref{fig:fuvebv}) using Planck reddening. We found relations of E(B - V) = $ 0.01124 + 1.119 \times 10^{-4} \times (FUV - 250)$ over the much larger area we observe in the NGP and E(B - V) = $ 0.01288 + 4.6841 \times 10^{-5} \times (FUV - 212)$ in the SGP. As \citet{Boissier2015} point out, the FUV emission is dependent on the geometry of the stars and the dust and care has to be taken when using the \galex\ data to predict extinction over the sky. 

\section{Modeling Milky Way Radiation}

\begin{figure}
\includegraphics[scale=0.4]{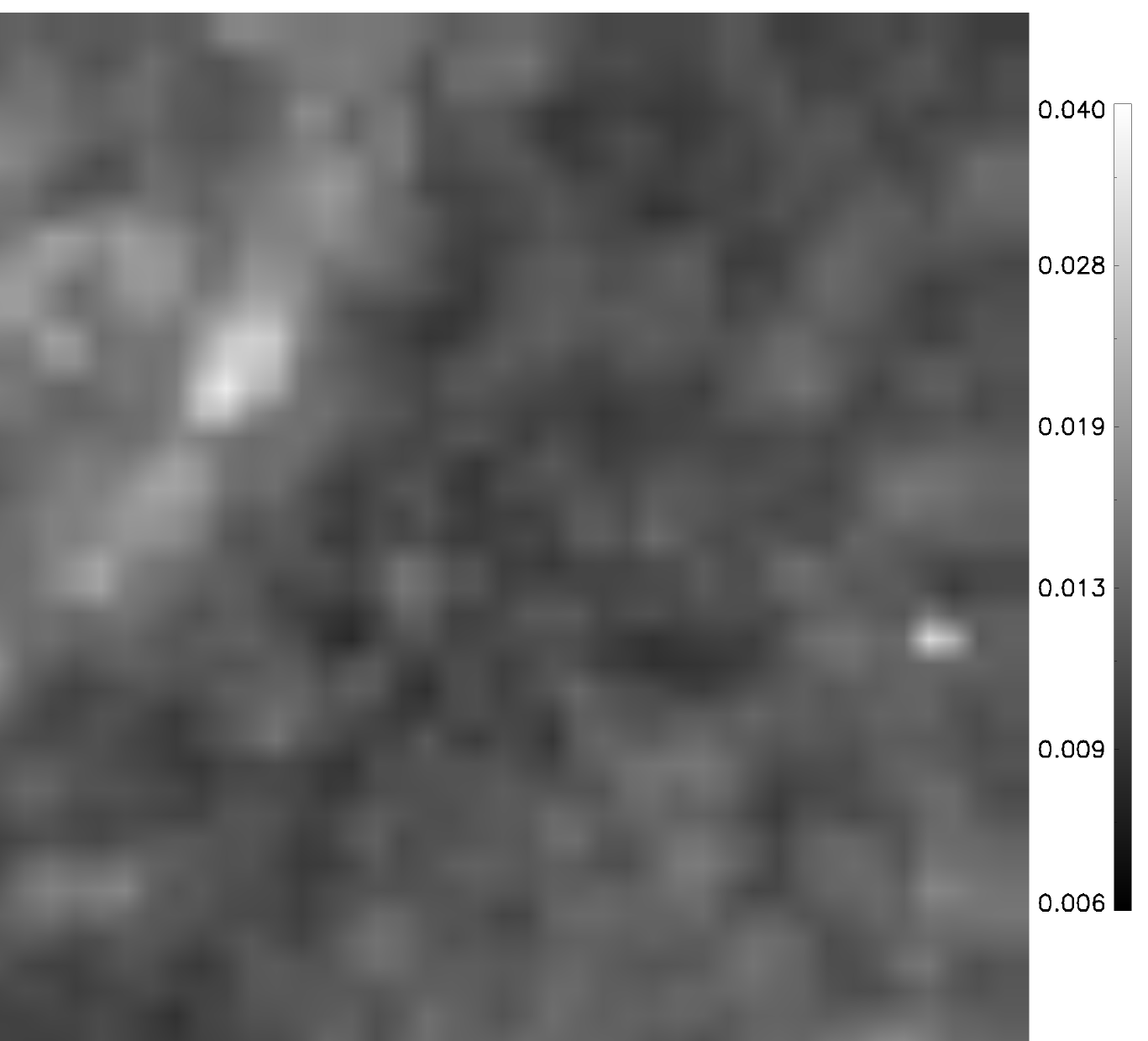}
\includegraphics[scale=0.4]{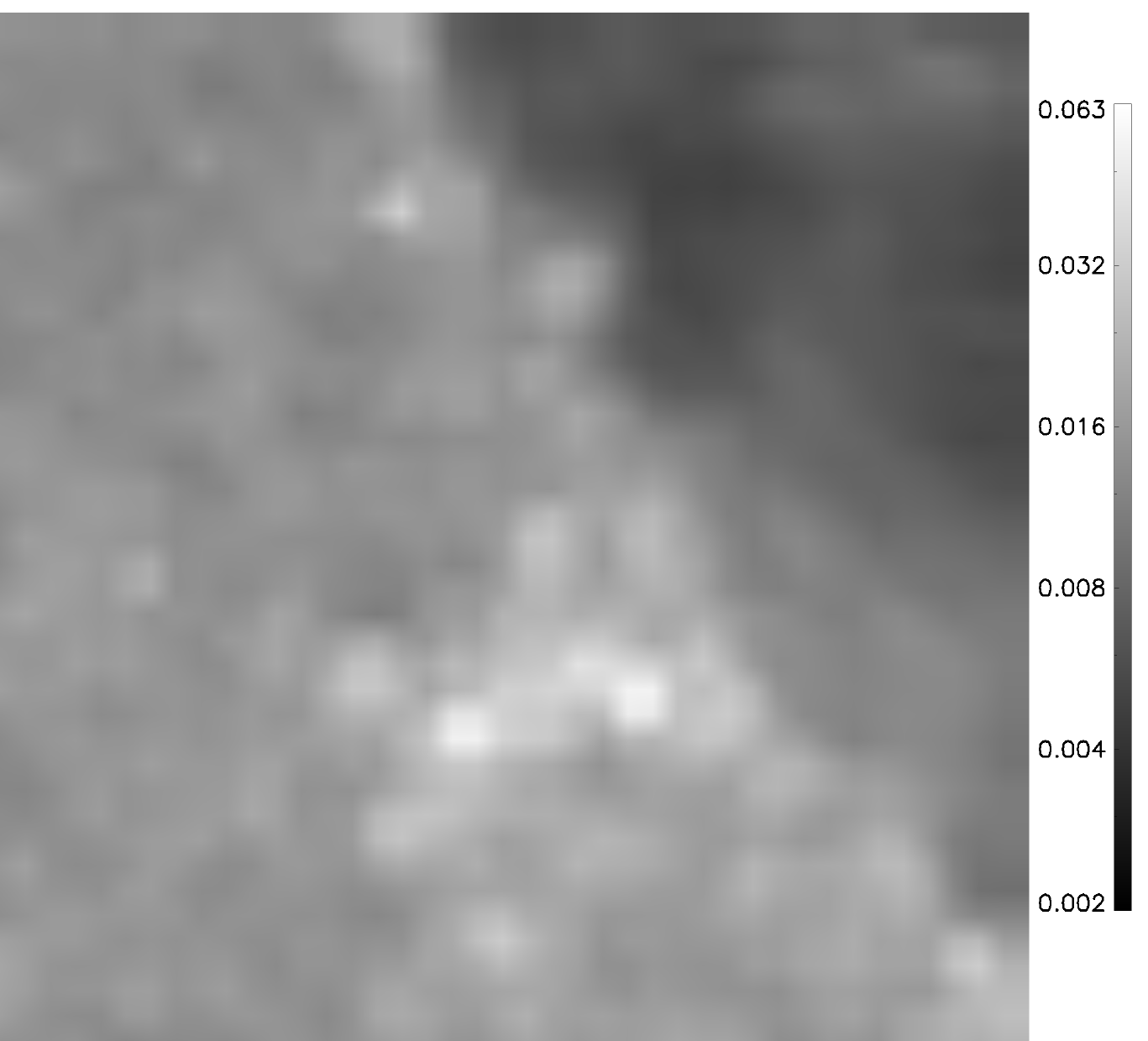}
\caption{Modeled dust distribution (E(B - V)) at the NGP (top) and SGP (bottom). To be compared with the 100 \micron\ plots in Figs. \ref{fig:ngp_images} and \ref{fig:sgp_images}.}
\label{fig:model_dust}
\end{figure}

\begin{figure}
\includegraphics[width=3.2in]{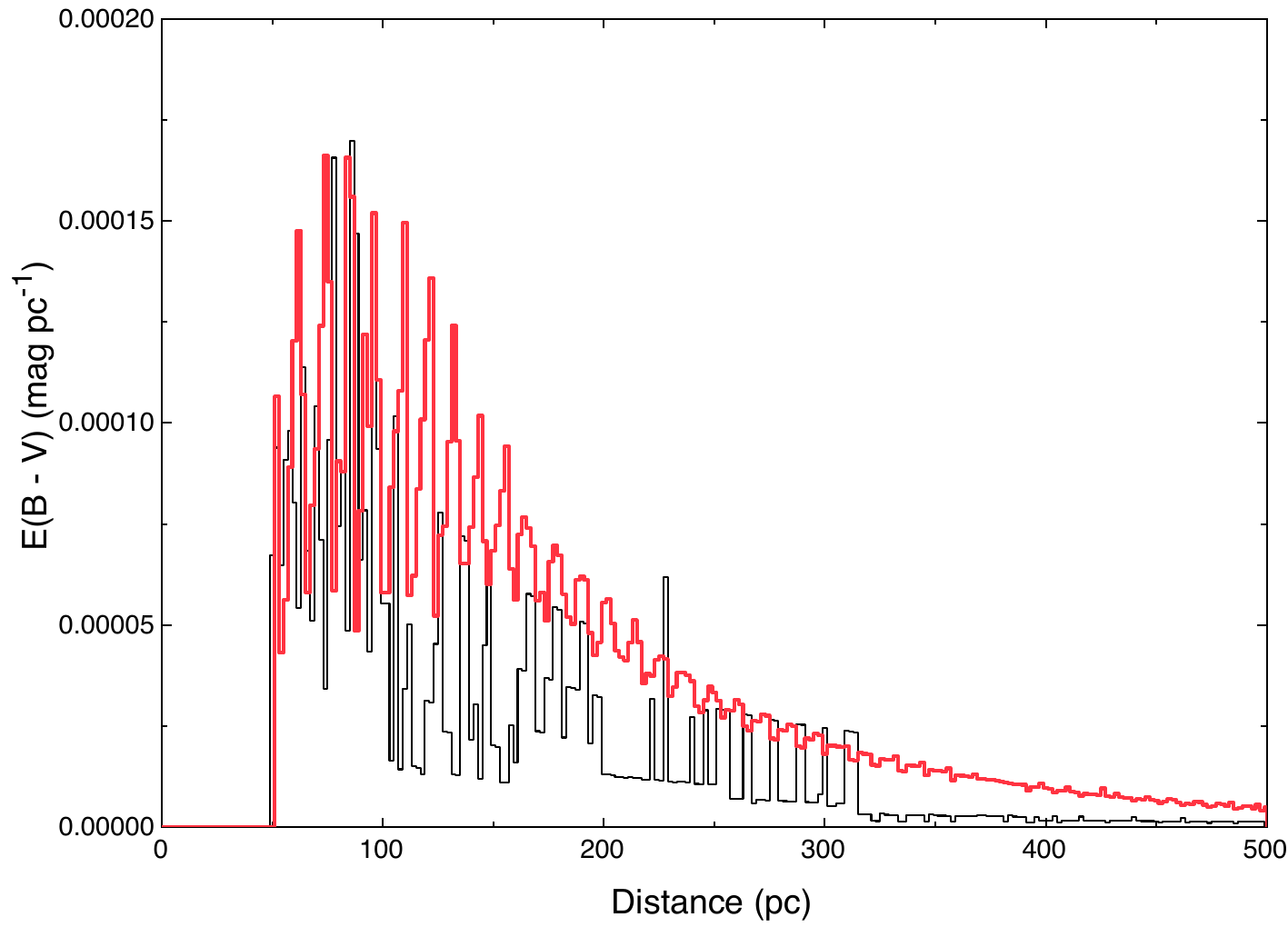}
\caption{Modeled extinction as a function of distance for the NGP and SGP (red line).}
\label{fig:model_dust_dist}
\end{figure}

Most of the DGL at low Galactic latitudes is unequivocally due to the scattering of the light of hot stars from interstellar dust and we have applied the model developed by \citet{Murthy2016} to predict the amount of Galactic dust-scattered radiation in the polar regions. This model uses a Monte Carlo process to track photons emitted from stars with location and spectral type from the Hipparcos catalog \citep{Perryman1997} and stellar spectra from \citet{Castelli2004}. The dust was taken from the 3-dimensional extinction map derived from PanSTARRS data by \citet{Green2015} with an angular resolution of about $14'$ at the poles. The gaps in the \citet{Green2015} map were filled using the reddening map given by \citet{Schlegel1998} with a scale height of 125 pc \citep{Marshall2006}. Our modeled dust distribution is shown in Fig. \ref{fig:model_dust} for both poles and is similar to the IR maps shown in Fig. \ref{fig:ngp_images} and Fig. \ref{fig:sgp_images}, respectively. The distribution of the extinction with distance (along a specific line of sight) is shown in Fig. \ref{fig:model_dust_dist} and is consistent with a scale height of 125 pc \citep{Marshall2006} and a cavity of about 50 pc radius around the Sun \citep{Welsh2010}. We assumed the scattering function of \citet{Henyey1941} with the albedo ({\it a}) and phase function asymmetry factor ($g = < cos \theta > $) as free parameters.

The dust at both poles has been extensively investigated through polarization measurements \citep{Markkanen1979, Berdyugin1995, Berdyugin1997, Berdyugin2000, Berdyugin2001, Berdyugin2002, Berdyugin2016, Berdyugin2004, Berdyugin2011, Berdyugin2014}. The polarization in the NGP was divided into two regions: Area I and Area II \citep{Markkanen1979}, approximately corresponding to with the 100 \micron\ surface brightness and the polarization being larger in Area II. The overall extinction in both poles is low with minimum values close to zero \citep{Fong1987, McFadzean1983}, except for limited areas where clouds are seen in the IR maps with peak values of E(B - V) from 0.02 -- 0.04 \citep{Berdyugin2011}. \citet{Berdyugin2014} found that the polarization was correlated with the IR maps with the caveat that the polarization maps probed the dust to a distance of about 400 pc while the IR emission measured the dust along the entire line of sight. \citet{Berdyugin2016} note that there may be some dusty structures extending to high positive latitudes within Area I, as suggested by the distribution of dark and molecular clouds, in addition to the diffuse dust. In general, we find that our dust model is in agreement with the polarization observations.

\begin{figure*}
\includegraphics[scale=0.4]{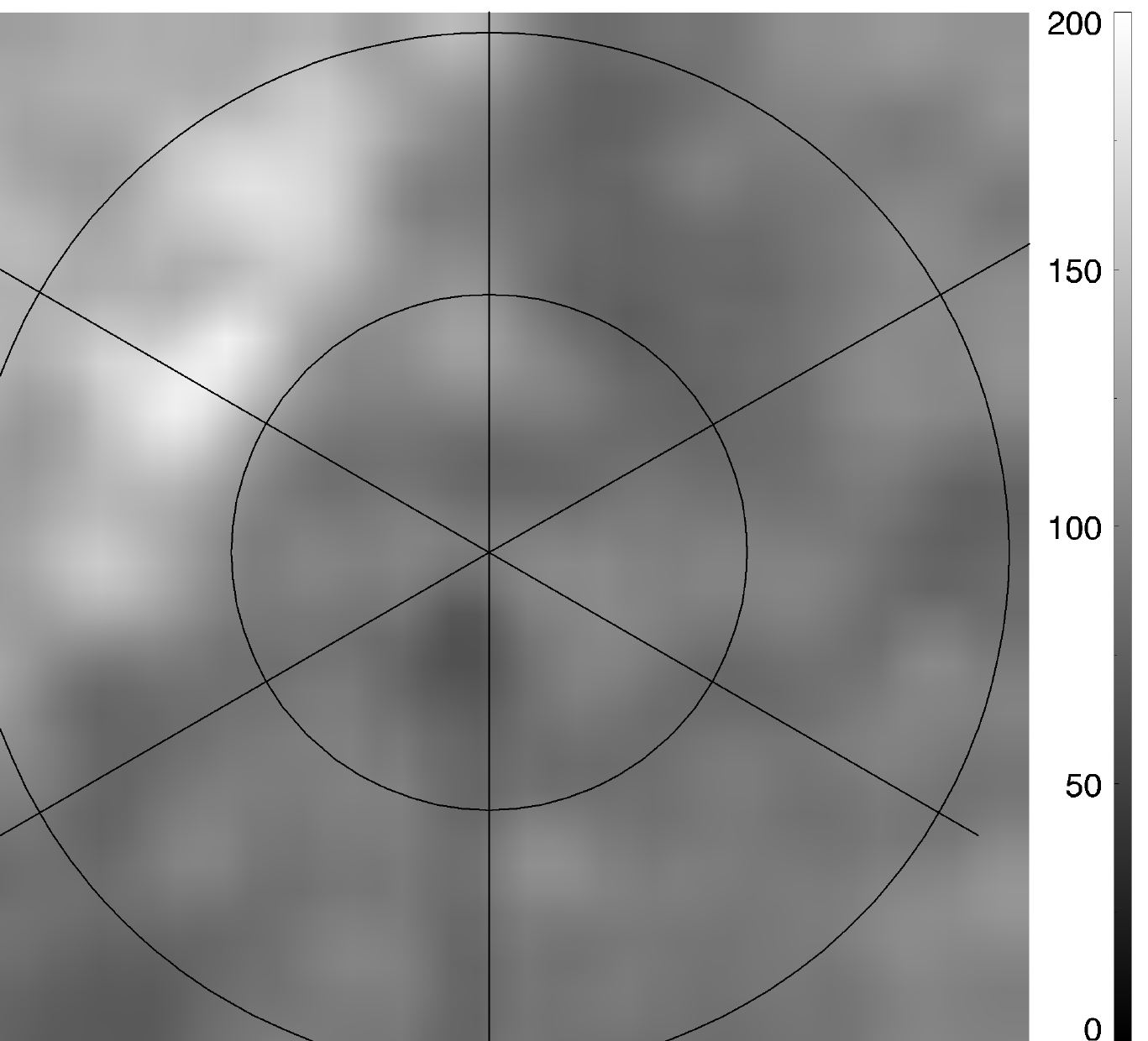}
\hspace{0.5cm}
\includegraphics[scale=0.4]{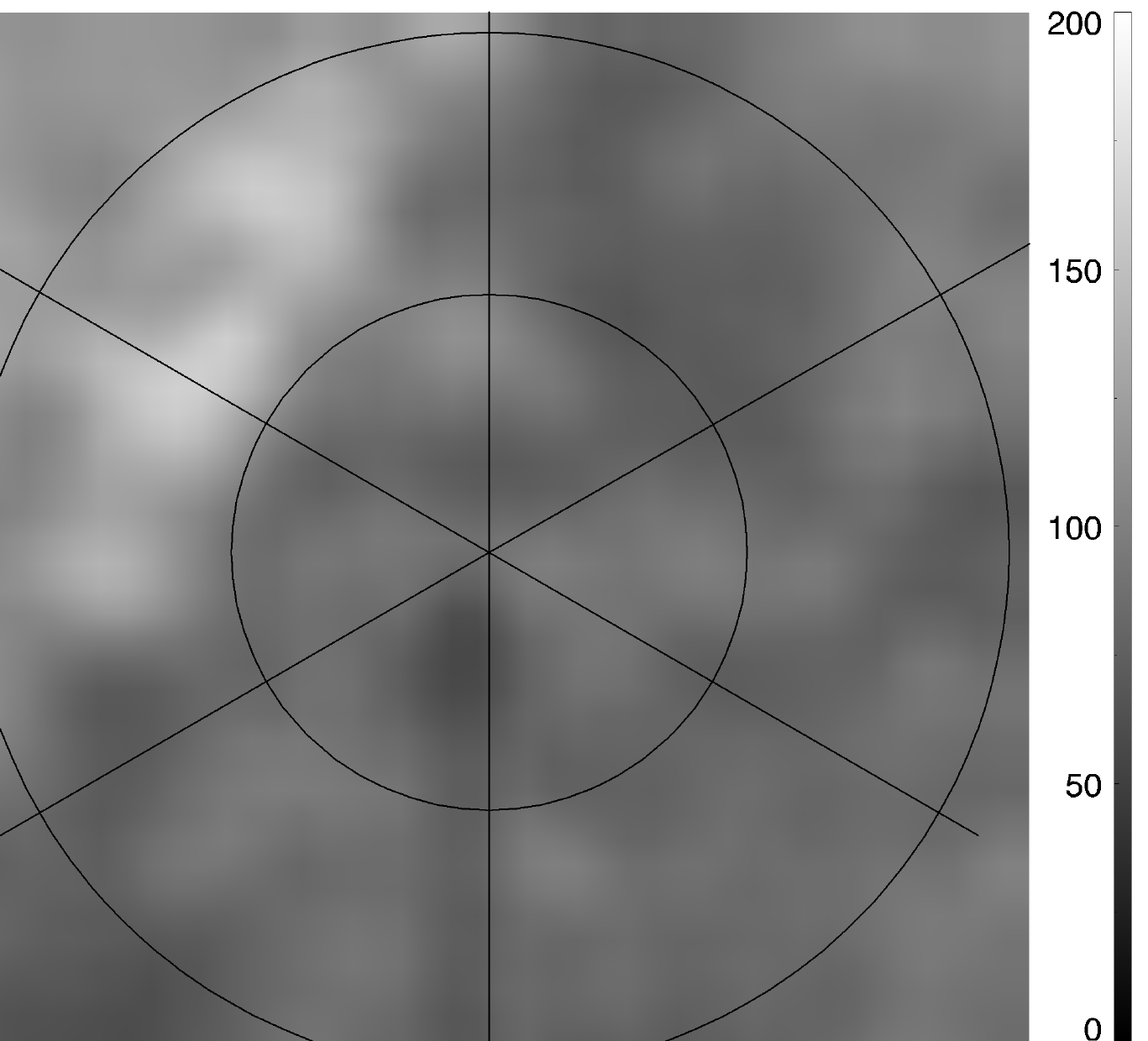}
\caption{Modeled surface brightness for the NGP in FUV (a=0.4, g=0.6) and NUV (a=0.4, g=0.5) at a resolution of $30'$. The maps are in photon units. The NGP is at the center with lines of latitude at $80$ and $85^{\circ}$ and lines of longitude every $60^{\circ}$ starting from $0^{\circ}$ at the top increasing clockwise.}
\label{fig:model_ngp}
\end{figure*}

\begin{figure*}
\includegraphics[scale=0.4]{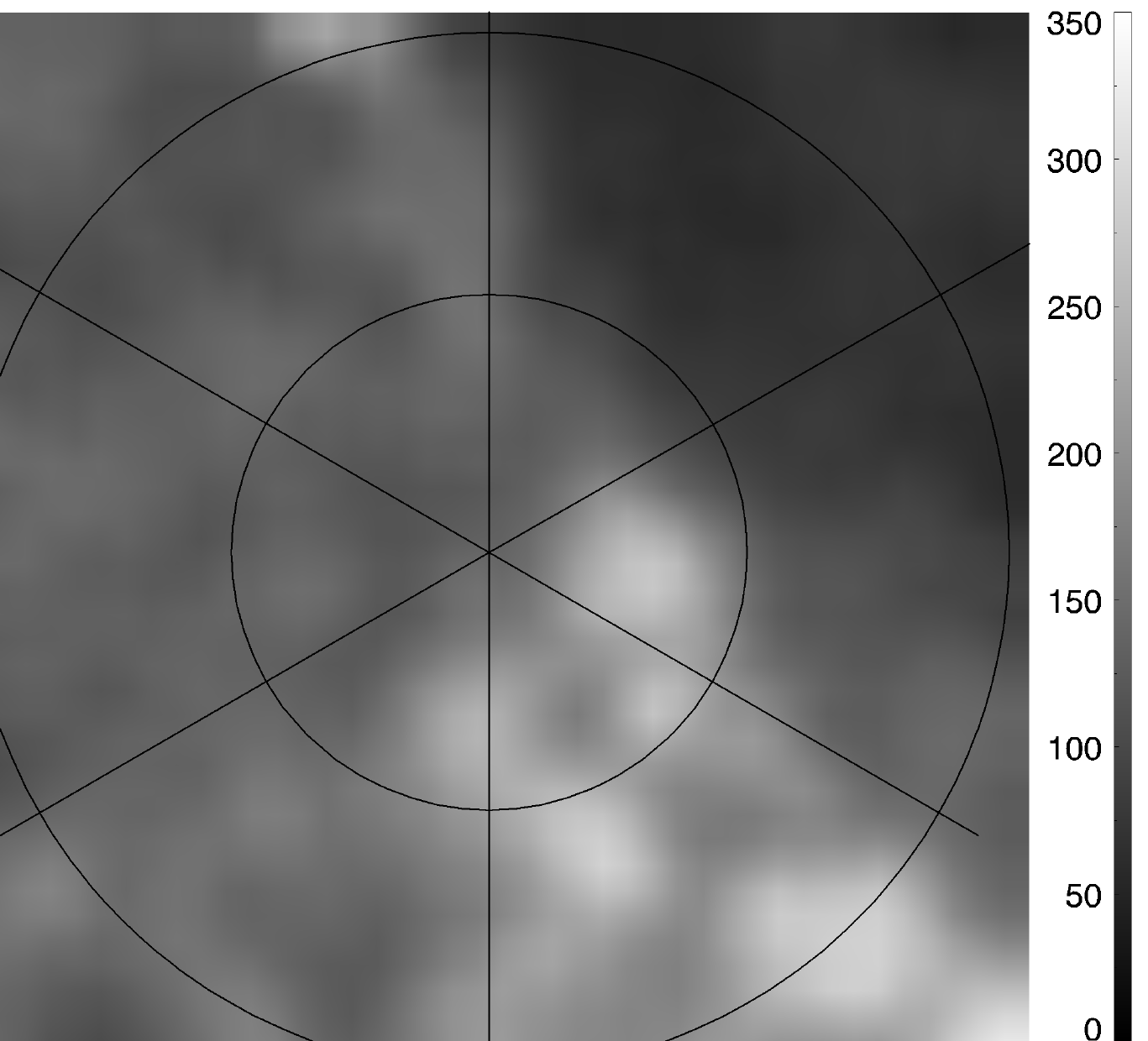}
\hspace{0.5cm}
\includegraphics[scale=0.4]{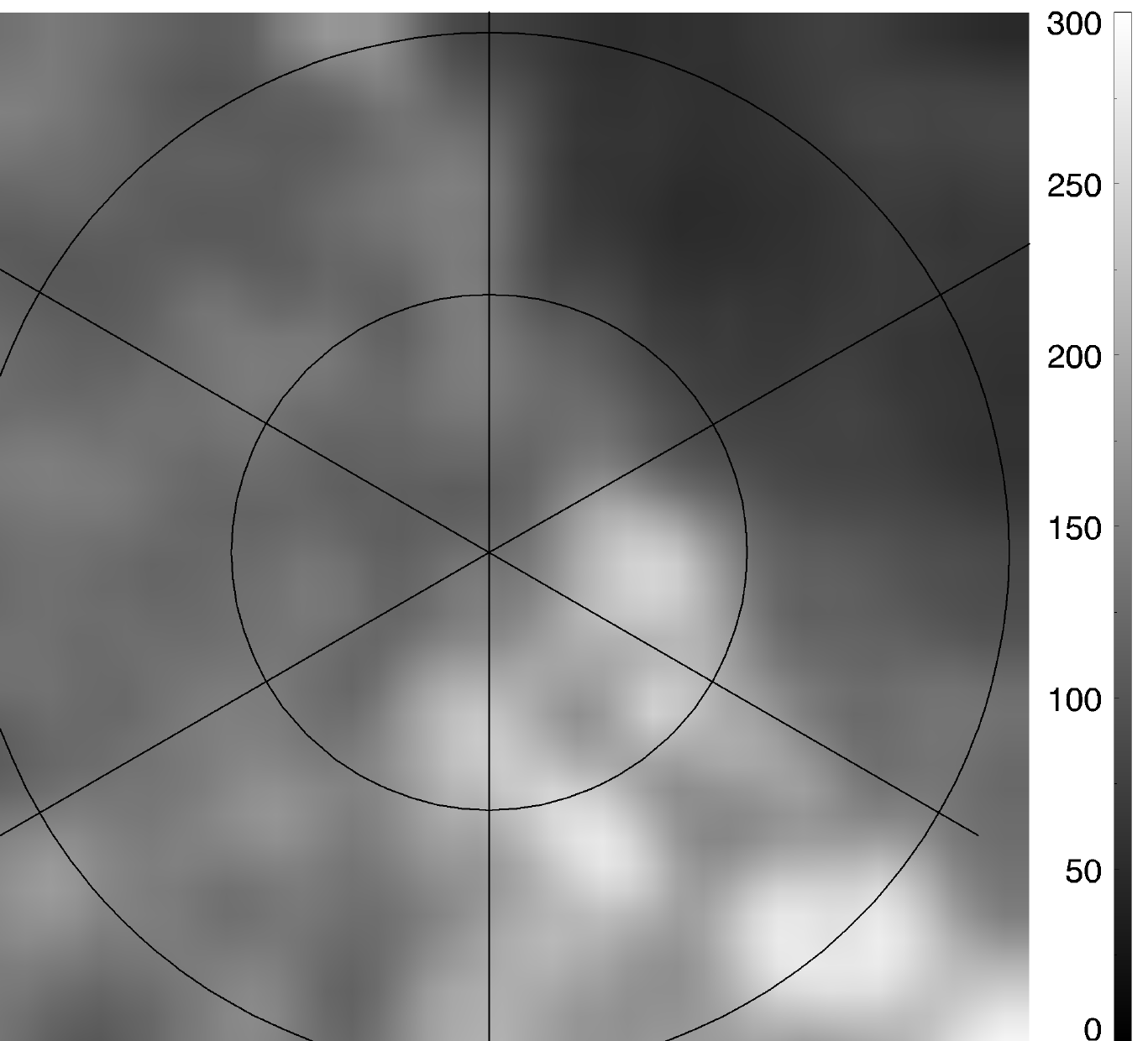}
\caption{Modeled surface brightness for the SGP in FUV (a=0.4, g=0.6) and NUV (a=0.4, g=0.5) at a resolution of $30'$. The maps are in photon units. The SGP is at the center with lines of latitude at $-80$ and $-85^{\circ}$ and lines of longitude every $60^{\circ}$ starting from $0^{\circ}$ at the top increasing anti-clockwise.}
\label{fig:model_sgp}
\end{figure*}

\begin{figure}
\includegraphics[trim = 1.5cm 12cm 1.5cm 3.5cm, clip, scale=0.5]{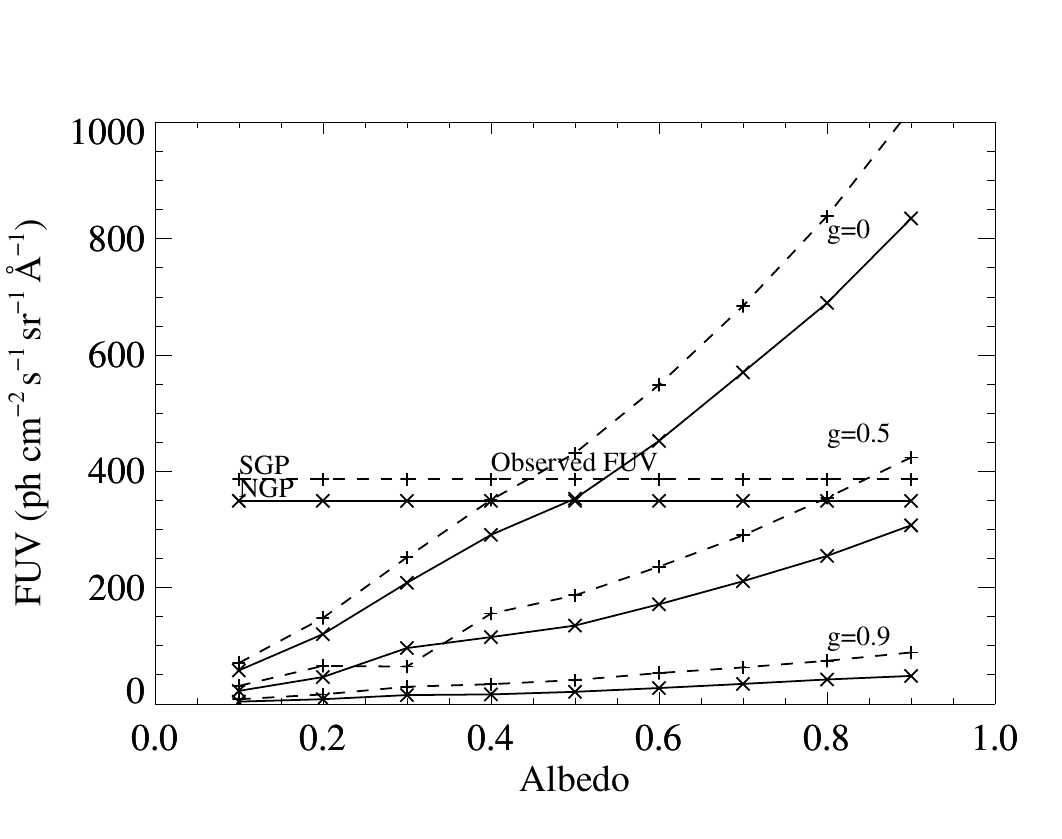}
\includegraphics[trim = 1.5cm 12cm 1.5cm 3.5cm, clip, scale=0.5]{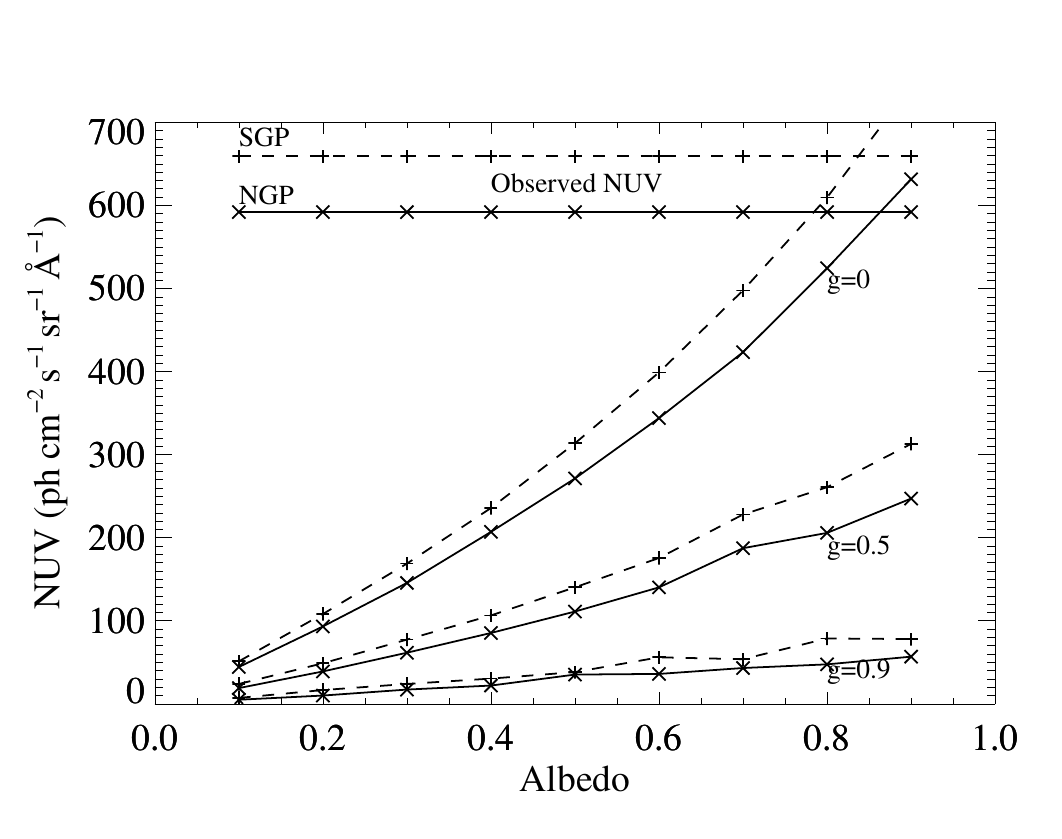}
\caption{Modeled surface brightness for the NGP (solid line) and the SGP (dashed line) fall short of the observed surface brightness in both the FUV and the NUV. The surface brightness is averaged over the entire region in there plots.}
\label{fig:dust_total}
\end{figure}

\begin{figure}
\includegraphics[trim = 1.5cm 12cm 1.5cm 3.5cm, clip, scale=0.5]{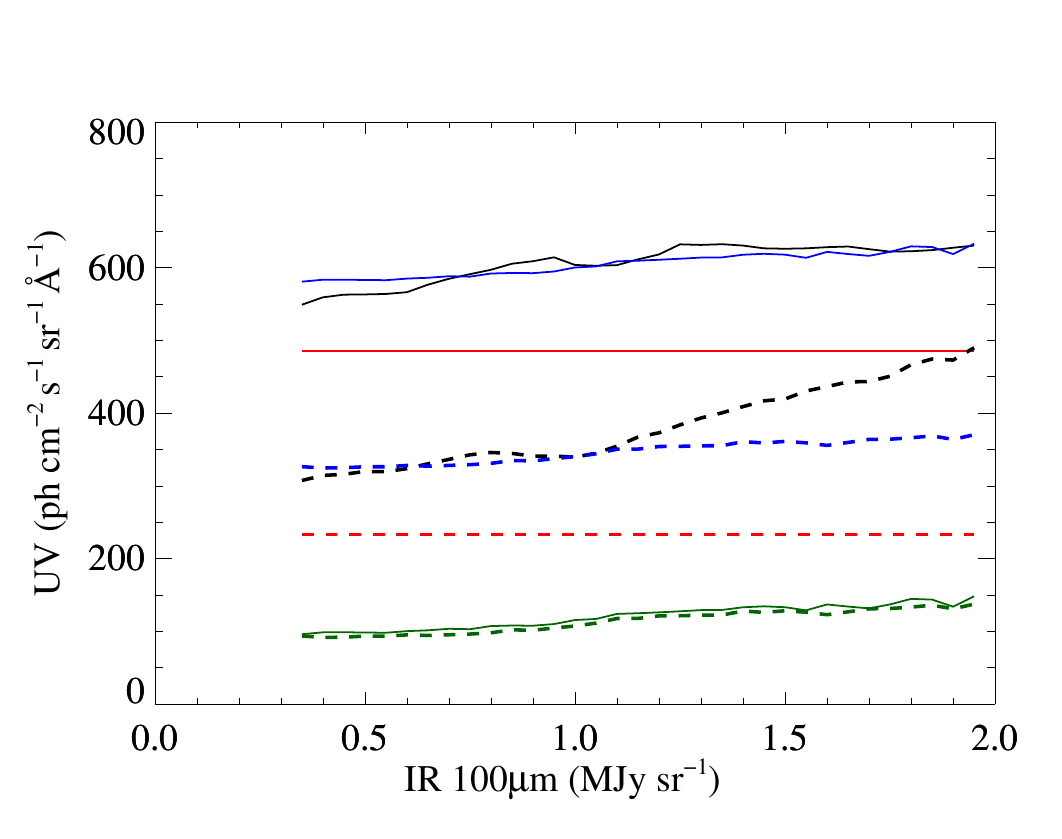}
\caption{FUV (dashed lines) and NUV (solid lines) modeled surface brightness (green line) plotted against the IR 100 \micron\ surface brightness with the observed background (black line) and the offset (red line) for the NGP. The blue lines show the models with the offsets of 233 \photu\ in the FUV and 485 \photu\ in the NUV. FUV model is for a=0.4 and g=0.6 and NUV model is for a=0.4 and g=0.5.}
\label{fig:model_uv_ngp}
\end{figure}

\begin{figure}
\includegraphics[trim = 1.5cm 12cm 1.5cm 3.5cm, clip, scale=0.5]{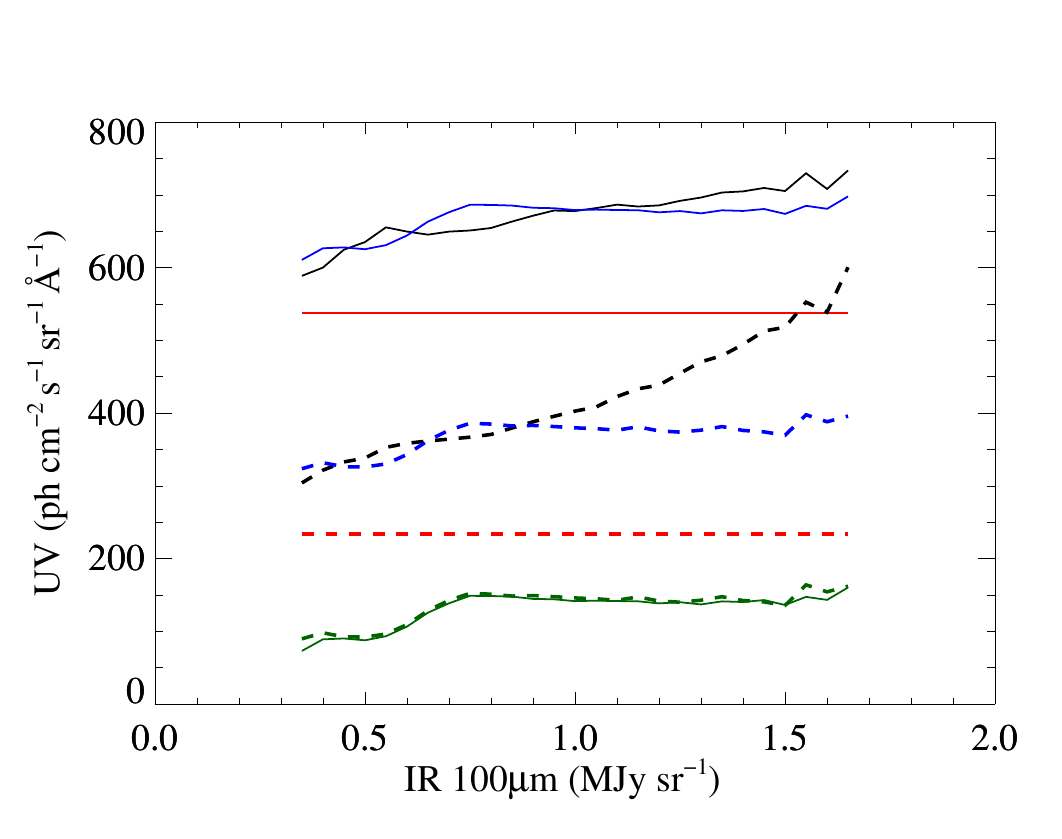}
\caption{FUV (dashed lines) and NUV (solid lines) modeled surface brightness (green line) plotted against the IR 100 \micron\ surface brightness with the observed background (black line) and the offset (red line) for the SGP. The blue line shows the models with the offsets of 234 \photu\ in the FUV and 538 \photu\ in the NUV. FUV model is for a=0.4 and g=0.6 and NUV model is for a=0.4 and g=0.5.}
\label{fig:model_uv_sgp}
\end{figure}

We have run our scattering model for a range of optical constants with representative results shown in Fig. \ref{fig:model_ngp} and \ref{fig:model_sgp}. The major dust features are clearly visible in the models but the brightness is much less than that observed unless the grains scatter isotropically (Fig. \ref{fig:dust_total}). Because most of the photons at the poles come from Galactic plane stars \citep{Jura1979}, the earliest papers did indeed find that $g = 0$. It is now generally accepted \citep{Draine2003} that the optical constants are close to $a = 0.4; g = 0.6$ in the FUV and $a = 0.4; g = 0.5$ in the NUV and we have used those models to fit the observed emission at each pole. There is too much noise in both the models and the data to compare on a pixel-by-pixel level and we have rather integrated both as a function of the 100 \micron\ values from \citet{Schlegel1998} in Figures \ref{fig:model_uv_ngp} and \ref{fig:model_uv_sgp}. 

The fit is good in both poles and both bands with best-fit offsets of 233 and 234 \photu\ in the FUV in the NGP and SGP, respectively and offsets of 485 and 538 \photu\ in the NUV in the NGP and SGP, respectively. These are not far different from the zero-point offsets in Table \ref{tab:back_limits}. We had previously noted the inflection point in the FUV-IR correlation at 1.08 MJy sr$^{-1}$ in the NGP; a comparison with the models shows that it is present in both poles in the FUV. As discussed above, this may be due to fluorescence from the Werner bands of molecular hydrogen.

\section{Light from dark matter?}

The continued presence of this unexplained excess in the diffuse background prompts us to briefly consider possible connections to nonstandard physics. Leading particle dark-matter candidates such as supersymmetric WIMPs or axions produce photons by annihilation or decay, but not at UV energies \citep{Henry2015}. Another possibility is offered by primordial black holes (PBHs), which emit Hawking radiation with an approximately blackbody spectrum peaking at the characteristic energy $E=\hbar c^3/(8\pi G M)$ for PBHs of mass $M$. Thus a background with $E=7$~eV (midway between our FUV and NUV energies) might be associated with PBHs of characteristic mass $M\approx 2 \times 10^{21}$~g. This value coincides with one of three narrow remaining theoretically allowed PBH mass windows \citep{Carr2016}, a so far unremarked coincidence that we find intriguing enough to explore briefly here. A plausible production mechanism for PBHs with masses close to this range has been identified by \citet{Espinosa2017}. The question is whether PBHs of this kind could contribute significantly to the unexplained excess identified above, whose bolometric intensity $Q_u=4\pi I_{\lambda}\lambda \approx 5\times 10^{-5}$~erg~s$^{-1}$~cm$^{-2}$ with $I_{\lambda}\approx$~180~photon units at $\lambda\approx 2000$~\AA.

PBH luminosity is very low, $L<\sim 2\times10^{-55}L_{\odot}(M/M_{\odot})^{-2}\approx 6\times 10^7$~erg~s$^{-1}$ \citep{Overduin2008}. If these PBHs make up the cold dark matter in the halo of the Milky Way, then their local density $\rho\approx 0.008 M_{\odot}$~pc$^{-3}$ \citep{Bovy2012}. If they are distributed uniformly, then the nearest one is located at a distance $\bar{r}=(\rho/M)^{-1/3}\approx$~100 AU. Its intensity  $Q=L/(4\pi\bar{r}^2)\approx 2\times10^{-24}$~erg~s$^{-1}$~cm$^{-2}$ as seen by us is far too low to account for $Q_u$. Alternatively, the total number of PBHs in the halo is $N=M_h/M\approx 1\times 10^{24}$ where $M_h\approx 1\times 10^{12}M_{\odot}$ \citep{Xue2008}. If these are clustered near the Galactic center at $R=8$~kpc, then the halo intensity $Q_h=N L/(4\pi R^2)\approx2\times 10^{-19}$~erg~s$^{-1}$~cm$^{-2}$. This is still 15 orders of magnitude too small. More realistically, if the PBH halo extends beyond the Sun and can be regarded as approximately uniform in the solar vicinity, then $Q_h={\cal L}R\approx7\times 10^{-17}$~erg~s$^{-1}$~cm$^{-2}$ where luminosity density ${\cal L}=L\rho/M\approx 2\times 10^{-33}$~erg~s$^{-1}$~cm$^{-3}$. This still falls short of $Q_u$ by 12 orders of magnitude, a discrepancy that cannot plausibly be attributed to non-uniformity in the PBH distribution. We infer that PBHs are not likely to contribute significantly to the astrophysical background, a conclusion reinforced by others \citep{Frampton2016}. The failure of this explanation, of course, only deepens the mystery.

\section{Conclusions}

We have used \galex\ data to study the diffuse ultraviolet background at both the North and South Galactic poles with two primary results:
\begin{enumerate}
\item There is an excess emission (over the DGL and the EBL) of 120 -- 180 \photu\ in the FUV and 300 -- 400 \photu\ in the NUV. Offsets in the UV emission have always been observed at the poles (Table \ref{tab:pole_obs}) but it has not been apparent how to attribute it to the different contributors. Although we do not know its origin, we can affirm that the excess emission cannot be accounted for by current models of the DGL and EBL.

\item We find that there is a change in the FUV-IR correlation at a 100 \micron\ surface brightness of 1.08 MJy sr$^{-1}$ (Fig. \ref{fig:model_uv_ngp} and \ref{fig:model_uv_sgp}). We believe that the most likely explanation for this is molecular hydrogen fluorescence indicating that self-shielding occurs at a column density of logN$_{H}$ = 20.2.
\end{enumerate}

We believe that the study of the Galactic poles will prove to be fruitful in differentiating between the Galactic and extragalactic (and terrestrial) components. Deep spectroscopy of the poles, including of cirrus features, would have been invaluable in separating the components but that seems unlikely in the near future with a dearth of UV missions expected.  In its absence, we will continue our in-depth study of diffuse emission with \galex.

\acknowledgments
We thank Prof. Berdyugin and Teerikorpi for clarifying the polarization results in the poles.
Part of this research has been supported by the Department of Science and Technology under Grant IR/S2/PU-006/2012. This research has made use of NASA's Astrophysics Data System Bibliographic Services. We have used the GnuDataLanguage (http://gnudatalanguage.sourceforge.net/index.php) for the analysis of this data. The data presented in this paper were obtained from the Mikulski Archive for Space Telescopes (MAST). STScI is operated by the Association of Universities for Research in Astronomy, Inc., under NASA contract NAS5-26555. Support for MAST for non-HST data is provided by the NASA Office of Space Science via grant NNX09AF08G and by other grants and contracts.

\bibliographystyle{yahapj}
\bibliography{references}

\end{document}